\documentclass[prb,twocolumn,superscriptaddress]{revtex4-1}

\usepackage{graphicx,color}
\usepackage{amsmath}
\usepackage{amssymb}
\usepackage{hyperref}
\usepackage[normalem]{ulem}

\newcommand{\ket}[1]{|{#1}\rangle}

 \newcommand\beq            {\begin{equation}}
 \newcommand\eeq           {\end{equation}}
 \newcommand\bwt         {\begin{widetext}}
 \newcommand\ewt         {\end{widetext}}

\begin{document}

\title{$PT$ invariant Weyl semimetals in gauge-symmetric systems}

\author{L. Lepori}

\affiliation{Dipartimento di Fisica e Astronomia,
Universit\`a di Padova, Via Marzolo 8, 35131 Padova, Italy.}
\affiliation{IPCMS (UMR 7504) and ISIS (UMR 7006), Universit\'{e} de Strasbourg and CNRS, Strasbourg, France.}

\author{I. C. Fulga}
\affiliation{Department of Condensed Matter Physics, Weizmann Institute of Science, Rehovot 76100, Israel.}

\author{A. Trombettoni}
\affiliation{CNR-IOM DEMOCRITOS Simulation Center, Via Bonomea 265, I-34136 Trieste, Italy.}
\affiliation{SISSA and INFN, Sezione di Trieste,
via Bonomea 265, I-34136 Trieste, Italy.}
 
\author{M. Burrello }
\email{for correspondence: michele.burrello@mpq.mpg.de}
\affiliation{Max-Planck-Institut f\"ur Quantenoptik, Hans-Kopfermann-Str. 1, 
D-85748 Garching, Germany.}

\begin{abstract}
Weyl semimetals typically appear in systems in which either 
time-reversal (${T}$) or inversion (${P}$) symmetry is broken. 
Here we show that in the presence of gauge potentials these topological states of matter can also arise in 
fermionic lattices preserving both ${T}$ and ${P}$. 
We analyze in detail the case of a cubic lattice model with $\pi$-fluxes, discussing 
the role of gauge symmetries in the formation of Weyl points and the difference between 
the physical and the canonical ${T}$ and ${P}$ symmetries. 
We examine the robustness of this $PT$ invariant Weyl semimetal phase against perturbations that remove the chiral sublattice symmetries and we discuss further generalizations. Finally, motivated by advances in ultracold atom experiments 
and by the possibility of using synthetic magnetic fields, we study the effect of random perturbations of the magnetic fluxes, which can be compared to a local disorder in realistic scenarios.
\end{abstract}

\maketitle

\section{Introduction} 
In recent years it has been understood that,
besides gapped phases of matter \cite{ludwig08,hasankane,zhang11}, gapless systems may also present topological properties. The main examples are topological metals and semimetals, whose band structure reproduces the main features of topological insulators and superconductors at their critical points. This is the case of the so-called Weyl semimetals \cite{wan11,balents11,burkov11}, whose recent experimental realizations  include tantalum arsenide \cite{hasan15c,lv15a,lv15b}, niobium arsenide \cite{hasan15d} and photonic crystals \cite{lu15}.

A crucial ingredient in obtaining Weyl semimetals is the breaking 
of  inversion (${ P}$) or time-reversal (${ T}$) 
symmetry \cite{balents11,dassarma14,jiang12,delplace,law2015}. 
In the absence of at least one of these symmetries 
it is possible to obtain zero-energy bulk modes with linear dispersion 
(Weyl nodes), which constitute monopoles for the Berry connection 
and are therefore protected against generic perturbations \cite{volovik2003}. 
The nodes lead to the appearance of zero-energy surface states (Fermi arcs), 
connecting the projections of Weyl points with opposite monopole charges on the surface Brillouin zone \cite{wan11} (see Fig. \ref{fig:weylsym}). 

If the Hamiltonian of the system is invariant under both ${ T}$ and ${ P}$, starting from a Weyl node at momentum $\vec{k}_0$ and applying ${ T}$ one obtains a node with the same chirality at $-\vec{k}_0$. The subsequent application of ${ P}$ creates a new Weyl point with opposite chirality in the original point $\vec{k}_0$. As such, one cannot isolate single monopoles of the Berry phase \cite{delplace} (see 
ref. \onlinecite{young12} for a similar analysis on Dirac semimetals). This mechanism, however, relies on the fact that both ${ T}$ and ${ P}$ commute with the translation operator.

Here we show that a different scenario can occur when gauge symmetries are present.  
In this case, the previous argument no longer holds because the Weyl points linked by $T$ and $P$ do not appear in $-\vec{k}_0$ and $\vec{k}_0$. In the presence of a gauge potential $\vec{A}$, we must distinguish between a momentum $\vec{k}$ and a physical momentum $\vec{k}+\vec{A}$: only the latter is inverted by $T$ and $P$.
Therefore, one must distinguish ``physical'' $T$ and $P$ symmetries, which are realized with the addition of space-dependent gauge transformations, and "canonical'' ${\cal T}$ and ${\cal P}$ symmetries, expressed in terms of translationally invariant operators, mapping the Hamiltonian to a different, but gauge-equivalent, form. This distinction is similar to the one between magnetic and canonical translation operators.

This observation is motivated by the recent experimental advances in ultracold atom setups. Several experiments showed that it is possible to obtain large artificial magnetic fluxes in optical lattices through laser-assisted tunnelings \cite{Aidelsburger2013,Miyake2013,bloch15}. This offers new possibilities to reach a Weyl semimetal regime based on cubic lattices subject to $\pi$-fluxes \cite{affleck,laughlin,Hasegawa,ketterle14}, not breaking the physical $T$ and $P$ symmetries. The experimental observation of Weyl points in these setups can be achieved through band mapping techniques, based, for example, on Bragg spectroscopy \cite{sengstock10}. An alternative detection procedure relies on the Berry curvature of the energy bands, recently obtained in two-dimensional setups through interferometric techniques \cite{duca15} and measurements of the band populations \cite{sengstock15}. These methods can generalize the results obtained for two-dimensional Dirac points \cite{tarruell12} to three space dimensions.

Our main conclusion is that, in the presence of gauge symmetries, one needs to break  ${\cal PT}$ to have Weyl points, but $PT$ can be preserved.
To derive our results, we examine first the role of gauge symmetry in the cubic lattice model with magnetic $\pi$-fluxes. It describes a three-dimensional generalization of the experimental results obtained in Refs.~\onlinecite{Aidelsburger2013,Miyake2013,bloch15}, and can be engineered by optical tools \cite{dalibard11,goldman13rev,ketterle14}. This model illustrates the simplest occurrence of Weyl semimetals in a ${PT}$ invariant system with gauge invariance. 
The robustness of this phase against the breaking of the chiral sublattice symmetry is shown.  
Later on, we study the effect of perturbations in the fluxes, which may characterize the  physical realization of this Weyl phase. Such perturbations, despite breaking of ${PT}$, do not spoil the Weyl physics for realistic system sizes and sufficiently small random variations. Finally we investigate the effects of a parabolic trapping potential on the Weyl cones and on the related Fermi arcs, showing that they survive, even if deformed, in this condition.

\section{Cubic lattice model with $\pi$ flux} 
Consider a tight-binding model of spinless fermions on a $\pi$-flux cubic lattice (we set the lattice constant to $a=1$).
Given a specific gauge, such a system
breaks either the canonical ${\cal P}$ \cite{ketterle14} or ${\cal T}$ 
\cite{LMT} symmetry, despite its physics being ${P}$ and ${T}$ invariant.
We work in the Hasegawa gauge \cite{Hasegawa}, with a vector potential:
\begin{equation} \label{api}
\vec{A}={\pi}\left(z-y,y-z,0\right),
\end{equation}
corresponding to a magnetic field 
$\vec B= \pi \left(1,1,1\right)$.
The tight-binding Hamiltonian of the system under the gauge potential \eqref{api} reads 
\begin{equation}
H(\{\theta\}) = -  \omega \, 
\sum_{\vec{r} \, , \, \hat{j}}   c^{\dagger}_{\vec{r} + \hat{j}} 
e^{i \theta_{j}\left(\vec{r}\right)} 
 c_{\vec{r} } + \ \mathrm{H.c.} \, ,
\label{peierls1a}
\end{equation}
where $c^\dag$ and $c$ are creation and annihilation operators for 
fermions on the lattice, $\omega$ is the hopping strength, $\hat{j} = \hat{x}, \hat{y}, \hat{z}$,
and $\theta_{j}\left(\vec{r}\right) = \int_{r_j}^{r_j+a} A_j \left(\vec{r}\right)  \mathrm{d} r_j$. Eq.~\eqref{api} leads to:
\begin{equation}\label{eq:abthetas}
 \theta_x = \pi\left(z-y\right),\quad \theta_y = \pi\left(y-z\right) + \pi/2\,, \quad \theta_z = 0 \, .
\end{equation}

This gauge potential is ${\cal P}$ invariant: an inversion transformation centered in a lattice site,
\begin{equation} \label{inversion}
x \to - x \,, \quad 
y \to - y \,, \quad 
z \to - z \,, 
\end{equation}
leaves all  phases $\theta_j$ unchanged, since $e^{i\pi(z-y)}=e^{-i\pi(z-y)}$. Therefore the physical and canonical inversion symmetries are equivalent with this gauge choice, $P={\cal P}$, and $H= P^\dag H P$.

The situation is different for ${\cal T}$. We define it
 as  complex conjugation of the hopping amplitudes, such that $\mathcal{T} H(\{\theta\}) \mathcal{T}^\dag=H(\{-{\theta}\})$; this relation implies that  
${\cal T}$ is broken with our gauge choice, since $e^{i\theta_y(y,z)}\neq e^{-i\theta_y(y,z)}$. The system is however $T$ invariant. Indeed, 
through a gauge transformation 
$\mathcal{U}_y c_{\vec{r}} \equiv e^{i\pi y} c_{\vec{r}}$, 
$H(\{-\theta\})$ 
is equivalent to $H(\{\theta\})$:
\begin{equation} \label{Tphys}
\mathcal{T} H\left(\{\theta\right\}) \mathcal{T}^\dag = \mathcal{U}_y^\dag H\left(\{\theta\}\right) \mathcal{U}_y\,.
\end{equation}

We conclude that the physical time-reversal transformation, 
which leaves the Hamiltonian in Eq.~\eqref{peierls1a} unchanged, 
takes the form $T=\mathcal{U}_y{\cal T}$, requiring a space-dependent 
gauge transformation to be implemented. In contrast, the space inversion ${\cal P}=P$ has a canonical, 
translational-invariant form with our gauge choice. 
A different gauge choice can lead to the opposite situation 
 where ${\cal T}=T$ and ${\cal P} \neq P$ (see ref. \onlinecite{ketterle14} and Appendix \ref{app:gauge}).
 
Our main point is that, no matter the gauge choice, 
the system is not ${\cal P}{\cal T}$ invariant, which is necessary to have Weyl points,
but actually it is invariant under the physical $PT$ symmetry, 
even though a space-dependent gauge transformation 
is required to make this invariance explicit. 

The effect of $\mathcal{U}_y$ is to flip the signs in the tunneling amplitudes of Eq. \eqref{peierls1a} along the $\hat{y}$ direction, and, given its local nature, can be applied independently on the boundary conditions of the system. This is a general feature of every gauge transformation.
For our particular gauge choice in Eq. \eqref{api} the physical $P$ and $T$ symmetries commute; the same happens with every gauge choice because, given a unitary transformation $\cal{W}$, one has $[\mathcal{W}^\dag P \mathcal{W},\mathcal{W}^\dag T \mathcal{W}]=\mathcal{W}^\dag[P,T]\mathcal{W}=0$.

\begin{figure}[t]
\includegraphics[width=0.95\columnwidth]{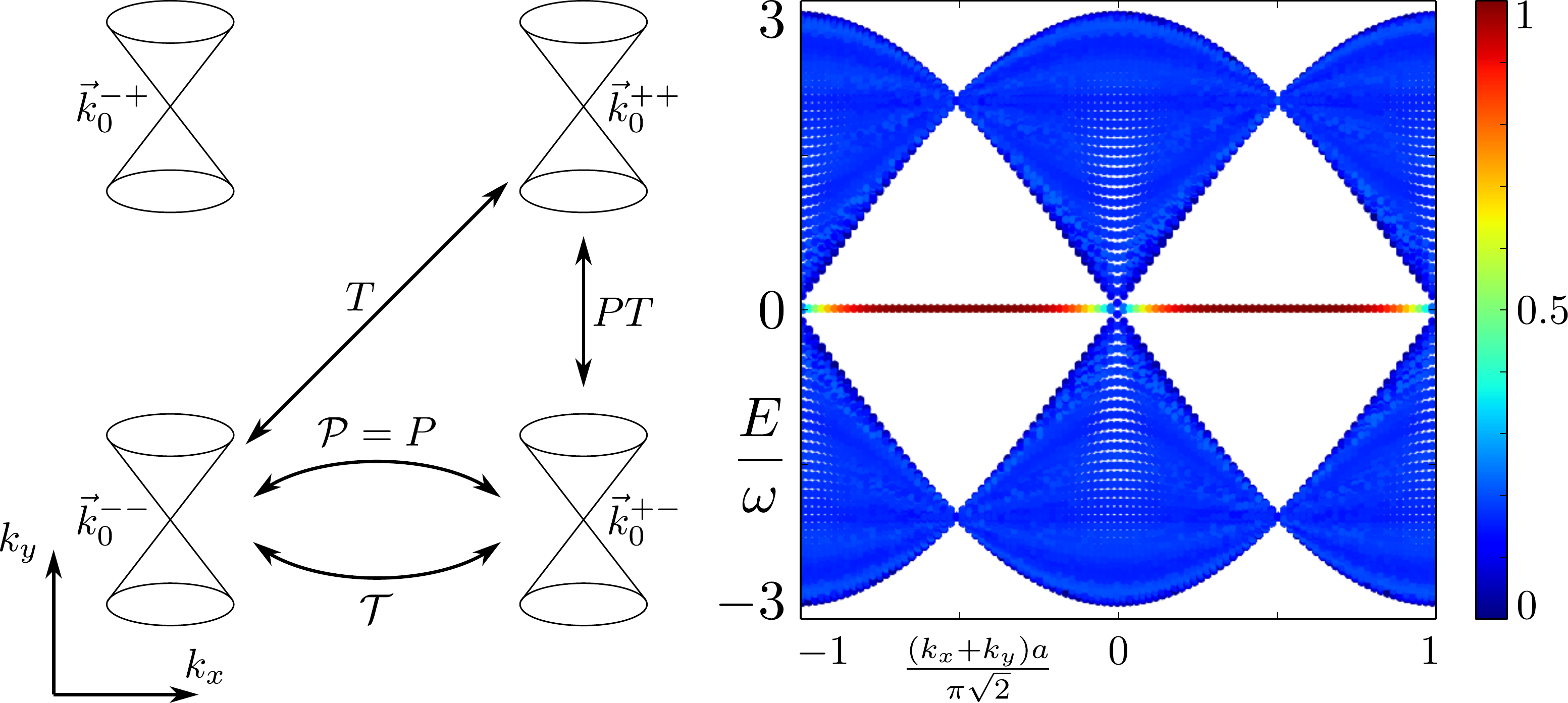}
\caption{Left: Effect of time-reversal and inversion symmetries on the Weyl cones of the Hamiltonian \eqref{hmag}. A  Weyl cone is mapped onto itself when applying the canonical ${\cal PT}$ symmetry. However, due to the gauge transformation ${\cal U}_y$, the physical $PT$ symmetry relates Weyl cones at different momenta. Right: Bandstructure in a diagonal slab geometry, with hard-wall boundaries in the $x-y$ direction and a thickness of 80 sites. The color scale indicates the amplitude of the eigenstates in the first and last eight sites from the boundaries. Weyl cones of opposite chirality are connected by zero-energy Fermi arcs (in red).} \label{fig:weylsym}
\end{figure}

Following refs. \onlinecite{Hasegawa, LMT}, we rewrite Eq. \eqref{peierls1a} in momentum space:
\beq \label{hmag}
H=  -  2\omega \, \big(\tau_x \cos k_x - \tau_y \cos k_y  + \tau_z \cos k_z \big),
\eeq
where $\tau_i$ are Pauli matrices and $\vec{k}$ belongs to the magnetic Brillouin zone (BZ). In the gauge \eqref{api}, we take $k_x,k_y\in\left[-\pi,\pi \right]$ and $k_z\in\left[0,\pi\right]$, and identify $(k_x,k_y,k_z)$ with $(k_x,k_y+\pi,k_z+\pi)$ \cite{Hasegawa}. The eigenstates of $\tau_x$ correspond to the even and odd sublattices in the $\hat{y}-\hat{z}$ plane. 

The energy bands of the system touch at the Weyl nodes $\vec{k}_0^{\pm,\pm}\equiv\left(\pm \pi/2,\pm\pi/2, \pi/2\right)$, having a chirality $\chi(\vec{k}_0^{\pm,\pm} ) =\prod_{j=x,y,z} \partial_{k_j} \cos k_j |_{\vec{k}=\vec{k}_0^{\pm,\pm}} $, such that the Weyl points at $\vec{k}_0^{\pm,\pm}$ are monopoles of charge $\chi$ for the Berry connection. 

In Eq.~\eqref{hmag}, ${T}$ does not take a canonical form,  allowing the formation of Weyl points. A Weyl point at $\vec{k}_0^{+,+}$ is mapped by $T=\mathcal{U}_y{\cal T}$ into a point  $ \vec{k}_0'=(-\pi/2,-\pi/2+\pi,-\pi/2)$ with the same chirality, corresponding to $\vec{k}_0^{-,-}$. ${\cal P}$  transforms $\vec{k}_0^{-,-}$ into a node with opposite chirality at $-\vec{k}_0^{-,-}=\vec{k}_0^{+,-}$. Therefore, the overall transformation ${\cal PT}\mathcal{U}_y$ relates inequivalent Weyl points with opposite chirality at \emph{different} momenta, $\vec{k}_0^{+,+}$ and $\vec{k}_0^{+,-}$, as shown in Fig.~\ref{fig:weylsym}.

The $PT$ invariant Weyl semimetal is a universal phase emerging from different lattice geometries with magnetic fluxes. In Appendix \ref{app:general} we present a family of Hamiltonians interpolating between the $\pi$-flux cubic lattice model and one defined by a stack of horizontal brickwall lattices, connected by vertical square plaquettes with $\pi$-fluxes. The models in this family are $PT$ invariant and display Weyl points. The definition of their physical symmetries relies, in general, on gauge transformations, and their effect on the Weyl cones can be analyzed following the general ideas presented above, 
leading to conclusions analogous to the ones depicted in the scheme in the left panel of Fig. \ref{fig:weylsym}a (see Appendix \ref{app:general}).

Weyl nodes have a topological origin, since they are monopoles of the Berry curvature for the energy bands. This feature induces stability against local noise \cite{wan11}. However, perturbations which couple two Weyl points with different momentum and opposite chirality cause them to annihilate pairwise. In the following we discuss the stability of the $PT$ invariant Weyl phase against perturbations breaking the chiral sublattice symmetry or consisting of small fluctuations of the fluxes in finite-size systems.

\subsection*{Breaking of the sublattice symmetry}
The Hamiltonian  in Eq. \eqref{peierls1a} is characterized by a chiral sublattice symmetry $c_{\vec{r}} \to (-1)^{x+y+z} c_{\vec{r}}$ mapping $H$ into $-H$, possibly broken by the introduction of staggering perturbations or onsite disorder. Below we show that this symmetry does not play any fundamental role in protecting the Weyl points: $PT$ invariant Weyl phases can exist also in systems without it.

We consider two staggering on-site potentials: $V_{ij} = V \, (-1)^{r_i-r_j}$ and $V_{xyz} = V \, (-1)^{x+y+z}$, both breaking the chiral symmetry.  In their presence,
the BZ is further halved along two directions. Correspondingly, the spectrum of Eq. \eqref{peierls1a} acquires additional degeneracies and Weyl points with different chiralities overlap in momentum space. A moderate perturbation $V_{ij}$ does not open a gap: it removes one of the additional degeneracies and it gaps only half of the Weyl points, leaving a pair of (degenerate) Weyl cones.
Therefore the system remains in the $PT$ invariant Weyl semimetal phase, despite the breaking of the sublattice symmetry.
The potential $V_{xyz}$, instead, immediately opens a bulk gap \cite{vishvanath10} by coupling Weyl nodes with different chiralities. However, this coupling is avoided when the Weyl points are displaced in momentum space by a term $V_{ij}$, stabilizing the Weyl phase against $V_{xyz}$ as long as $|V_{ij}|>|V_{xyz}|$. We conclude that, even in the absence of a chiral sublattice symmetry, there exists in general an extended Weyl semimetal phase invariant under the physical $P$ and $T$ symmetries, and stable against onsite perturbations not breaking them.

\section{Perturbation of the fluxes}  
A deviation of the fluxes from $\pi$ can couple separated points in the BZ; therefore it is important to study its effect on the Weyl phase,
crucially evaluating the range of errors in the fluxes still enabling the observation of the Weyl cones. We consider then the vector potential:
\begin{equation}
\label{eq:vpotchange}
 \vec{A} = \left( \beta z - \gamma y,  \pi y - \alpha z, 0\right), 
\end{equation}
corresponding to $\vec{B} = (\alpha, \beta, \gamma)$. This configuration implies  in general a breaking of $T$ and $P$, driving the system from the $PT$ invariant point to one with no symmetries. Indeed, both $T$ and $P$ invert all the magnetic fluxes, thus mapping the model into a gauge inequivalent system, unless $\alpha = \beta = \gamma =\pi$.

\begin{figure}[tb]
 \includegraphics[width=\columnwidth]{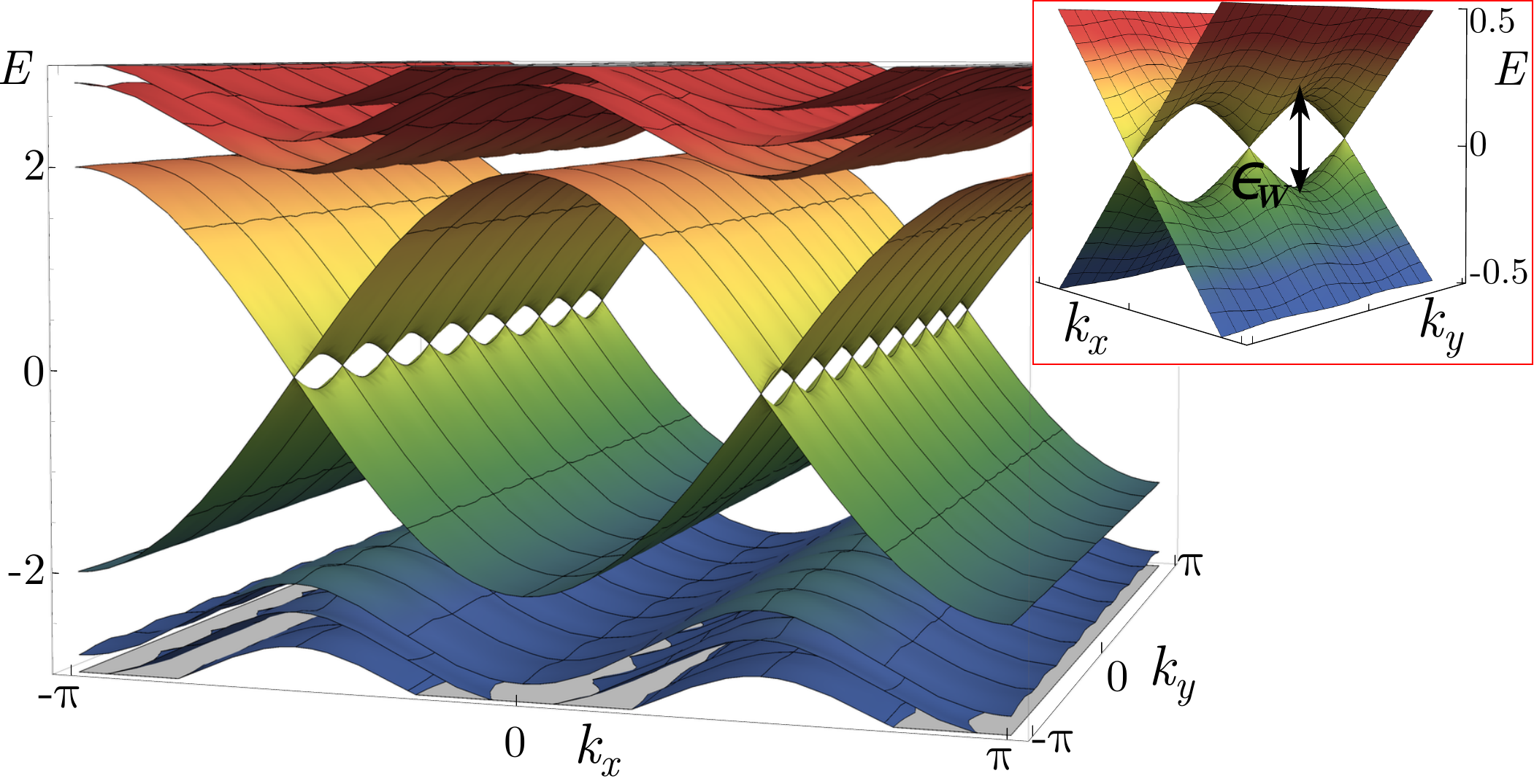}
 \caption{Spectrum for $\vec{B}=(3/4,1,1)\pi$ with $k_z=0$, displaying Weyl cones at zero energy. The inset shows the detail of a Weyl cone: the topological semimetal physics emerges clearly below an energy scale $\epsilon_W$, whereas above it the system has a metallic behavior.}
 \label{anis}
\end{figure}

We consider flux configurations $\phi=\pi (1-1/q_\phi)$, with $\phi=\alpha,\beta,\gamma$ and $q_\phi$ being a triplet of integers, leading to an increase in the magnetic unit cell, now whith a volume given by the least common multiple (LCM) of $q_\phi$: $V_{uc} = {\rm LCM}(q_\alpha,q_\beta,q_\gamma)$ \cite{notetras,landau}. We find that in the thermodynamic limit flux perturbations can have a drastic effect on the low-energy physics of the system, also due to the fractal nature of its spectrum \cite{koshino2001,koshino2003}. For instance, when $q_\alpha=q_\beta=q_\gamma>2$ the system is in a metallic state at half filling, with a two-dimensional Fermi surface.
A different case occurs for anisotropic configurations $\vec{B}=(\pi(1-1/q), \pi, \pi)$: for even $q$ the spectrum shows Weyl cones, whose density in the magnetic Brillouin zone increases with $q$. This  feature requires larger and larger system sizes to be seen for decreasing perturbations $1/q$. For example, in Fig. \ref{anis} we show the spectrum for $\alpha=3\pi/4$ and $\beta=\gamma=\pi$.  Interestingly, the physics of these Weyl points appears only at relatively low energies, characterized by a threshold $\epsilon_W$, then a sufficiently high energy resolution is required to detect the Weyl cones. For energies between  $\epsilon_W$ and the bandwidth, instead, the system is well approximated by a metallic behavior. 
For these anisotropic flux configurations the chiral symmetry constraints the Weyl points to be at zero energy, even in the presence of a simultaneous breaking of $T$ and $P$, differently from the standard behavior in other setups \cite{burkov12,soliacic13}.
Finally, for odd $q$ the low energy physics is described by a node-line semimetal \cite{schnyder15} with a one-dimensional Fermi surface, instead of a Weyl semimetal.

In ultracold atom experiments, lattice sizes are typically limited to tens of sites. Therefore, for small perturbations from $\alpha = \beta = \gamma =\pi$, the available system sizes are much smaller than the magnetic unit cell, $V \ll V_{uc}$. Below we show that, in this limit, the Weyl physics is robust to small variations of the fluxes. To examine these perturbations,  we compute the density of states (DOS) of a  cubic lattice with a finite linear dimension $L$, using a kernel polynomial approximation (see Appendix \ref{app:kernel}). The DOS is averaged over random flux configurations where $\alpha,\beta,\gamma$ are independent variables of the form  $\phi = \pi + \delta \phi$, and $\delta \phi$ is drawn randomly from the uniform distribution $[-U,U]$, with $U$ being the maximum perturbation. In this way, both the strength and the direction of  $\vec{B}$ are varied. The presence of a Weyl semimetal phase can be determined by examining the energy dependence of the average DOS, taking the form $\rho(E) \sim E^2$ for linearly dispersing states.

We investigated different values of $U$ up to $0.05$, for different system sizes up to $L=160$. Our results for the DOS share similarities with the physics of Weyl semimetals in the presence of onsite disorder  discussed in \cite{gurarie15,Pixley2016}. For relatively small values of $U$,
the zero-energy density $\rho(E=0)$ is different from zero 
and, in agreement with recent numerical  
results \cite{Pixley2016}, two 
energy thresholds emerge:  $\epsilon_1$ 
such that the average DOS 
is approximately constant for $|E| \leq \epsilon_1$, and $\epsilon_W$ 
such that $\rho(E) \propto E^2$ for $\epsilon_1 < |E| \leq 
\epsilon_W$, as expected for Weyl points.

The threshold $\epsilon_1$ signals in general the appearance of a diffusive behavior \cite{fradkin86,kobayashi14} at low energies.
Recent works \cite{Pixley2016,nandkishore14} suggest that this constant DOS can be associated with rare power-law localized states, although for flux perturbations its origin may be different. Independently on their nature, these states may hinder the observation of Weyl quasiparticles: the visibility of the Weyl phase depends in general on the width of the energy window $\epsilon_W-\epsilon_1$, which must be significantly large.

\begin{figure}
 \includegraphics[width=0.9\columnwidth]{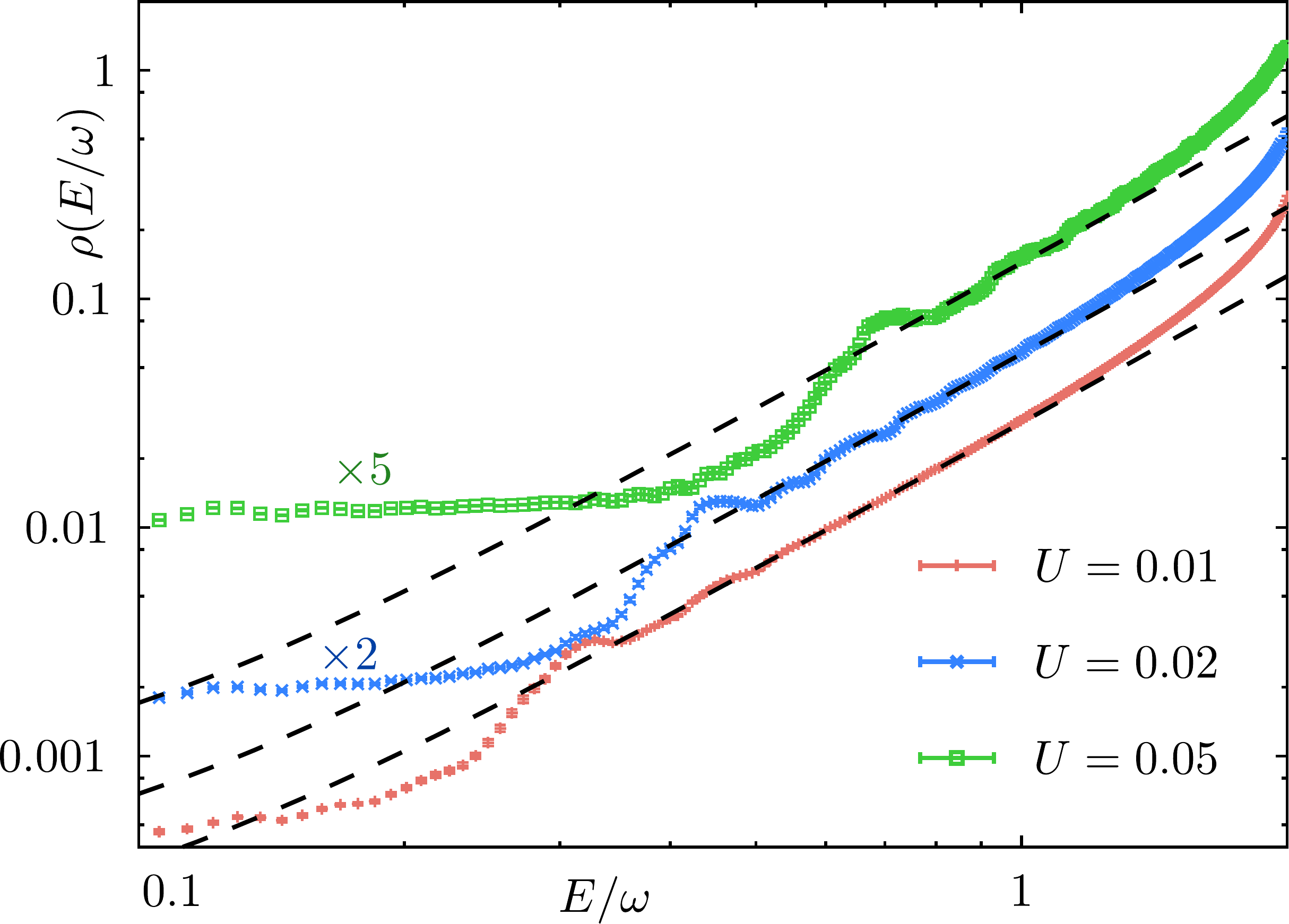}
 \caption{Double logarithmic plot of the density of states, as a function of energy for different values of $U$. The DOS is computed for a cube with  linear size of $160$ sites, and is averaged over $\sim150-200$ random flux realizations. For clarity, the green and blue curves are shifted vertically by multiplying with a constant factor (shown on the plot). 
The DOS close to the Fermi level is constant, indicating a diffusive metal behavior. In the intermediate energy range and for small values of $U$ we observe an energy scaling characteristic of Weyl semimetals (dashed lines), $\rho(E) - \rho(0) = a |E|^b$ with $b\approx 2$. Larger values of $U$ lead to an increase in the energy range associated with a diffusive metal, and a reduction of the region associated with a topological semimetal.
 }
 \label{dens}
\end{figure}

The average DOS for $L=160$ is plotted in 
Fig.~\ref{dens}. For $U=0.01$, we find $\epsilon_1 \approx 0.3 \, \omega$ and 
$\epsilon_W\approx \omega$. Close to the Fermi level the DOS takes an energy independent value $\rho(E/\omega=0)  \approx 10^{-3}$, whereas at intermediate energies 
we recover the scaling associated with a Weyl semimetal phase. 
Using the fit  $\rho(E) - \rho(0) = a |E|^b$ for $U=0.01$ inside 
the interval $[\epsilon_1,\epsilon_W]$ (we choose 
$0.52 < |E|/\omega < 0.75$) we obtain 
$a\simeq 0.029$ and $b\simeq 2.12$, with a reduced $\chi^2\simeq 1.02$. 
For smaller system sizes the qualitative behavior is very similar, even 
though the average DOS shows pronounced finite size effects, leading to a larger value of $\chi^2$. As $U$
is progressively increased, the energy $\epsilon_1$ associated 
with diffusive metal behavior also increases, while the topological 
semimetal region between $\epsilon_1$ and $\epsilon_W$ becomes smaller, eventually disappearing (as shown for $U=0.05$ in Fig.~\ref{dens}).

Our results indicate that the Weyl semimetal behavior is observable in ultracold 
atom implementations of the $\pi$-flux cubic lattice model, given
variations in the flux of $1\%-2\%$ for $L=160$ (a realistic estimate for setups of this kind \cite{Aidelsburger2013,Miyake2013,bloch15}) and 
generally $\delta\phi \lesssim \pi/L$.

\section{Effect of a confining potential} \label{app:trap}

Among the perturbations relevant for ultracold atom setups, which may have an impact on observing Weyl points 
and Fermi arcs, the presence of trapping potentials deserves particular attention.

Experimental setups in optical lattices usually rely on harmonic traps or hard-wall potentials obtained through optical box traps \cite{gaunt13} or light-intensity masks \cite{corman14,corman14b}, and, typically, they reach a system size of the order $L \approx 100$. We examine here the fate of the zero-energy modes \cite{wan11} in the presence of both these potentials.

Since the presence of a boundary along the $\hat{y}$ axis does not break the gauge invariance of Eq.~(2) of the main text, it does not spoil the distinction between canonical and physical symmetries, relying on the $\mathcal{U}_y$ gauge transformation. One consequence is that the same boundary can still host Fermi arcs, as in different geometries not involving the  $\hat{y}$ axis.
Keeping in mind this observation, we consider a diagonal slab geometry 
with hard wall boundaries located at $|x-y| = L$, 
and a harmonic trapping potential depending only on the distance from the surface. 
The single-particle Hamiltonian reads
\begin{equation}
 H_{\rm conf} = H + \frac{m}{2} \, \Omega^2 (x-y)^2 -\mu,
\end{equation}
with $H$ as in Eq. (2). For a small chemical potential, $0<\mu<2\omega$, a metallic phase appears at $x=y$, with the second energy band partially filled. In a local approximation, the  zero-energy Fermi surface corresponds to contours around the Weyl points.
Moving towards the surface, the harmonic trap reduces the local chemical potential. Therefore, 
the Fermi surface shrinks into the Weyl points, which are reached 
at a distance ${\cal R}$ such that $m \, \Omega^2 {\cal R}^2 = 2\mu$. 
Further increasing $|x-y|$, the system is driven again in a metallic phase, and, reaching the surface at $|x-y|=L$, zero-energy surface modes appear.

\begin{figure}[tb]
 \includegraphics[width=0.8\columnwidth]{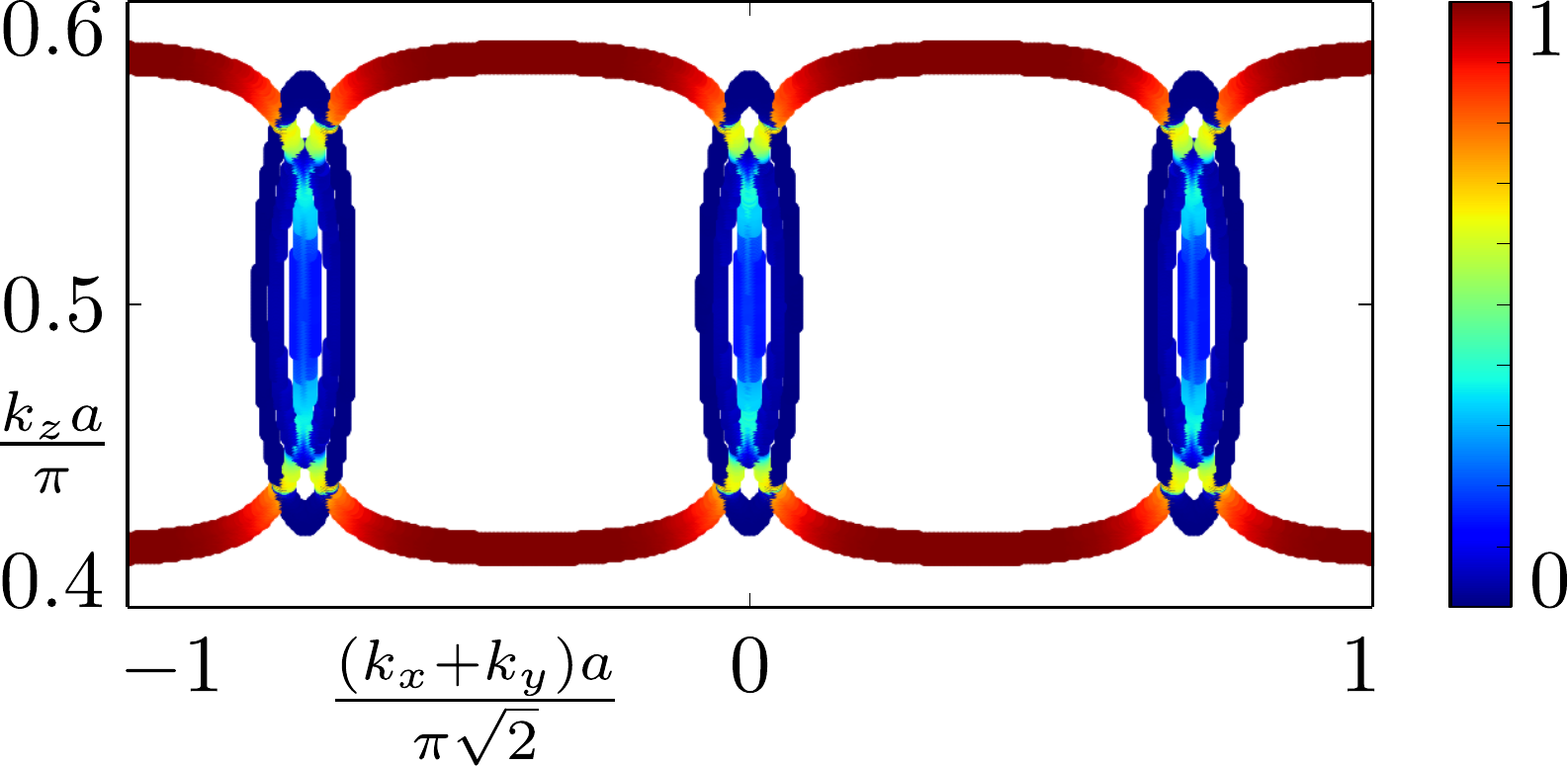}
 \caption{States with energy $|E|<0.02\omega$ in the presence of a harmonic trap in the surface BZ. The color depicts their localization close to the hard-wall boundary. Here $L=50$, $\mu=\omega/2$ and $\Omega$ is chosen  to have the potential energy ranging from $-\omega/2$ to $+\omega/2$ from the center towards the surface. The Weyl points become extended regions of zero-energy bulk states, connected by curved Fermi arcs.} 
\label{fermitrap}
\end{figure}

Figure~\ref{fermitrap} shows the states close to zero energy on the surface BZ in a diagonal slab geometry.
As expected, the zero-energy states of the full system include two regions 
of bulk states surrounding the projections of the Weyl points. Surface modes 
interpolate, with a curved shape, between the Fermi surfaces around the Weyl points. These Fermi arcs can be experimentally probed through a band mapping based on Bragg spectroscopy \cite{sengstock10}.

\section{Outlook} In fermionic systems with gauge invariance 
the physical time-reversal and inversion symmetries must be defined up to local gauge transformations. Such transformations have the non-trivial effect of translating states in momentum space, allowing to avoid the usual doubling of the zero-energy Weyl points present in systems with canonical time-reversal or inversion symmetries. 
Therefore, local gauge symmetries can be exploited 
to define a new kind of $PT$ invariant Weyl semimetal.
We discussed this phase for the simplest case of a cubic lattice model of spinless fermions with $\pi$-magnetic fluxes, 
showing that its Hamiltonian is invariant under a transformation 
$\mathcal{PTU} \equiv PT$, despite the presence of Weyl points in its spectrum. 
We emphasize, however, that the existence of the $PT$ invariant Weyl phase is not limited to spinless fermions with a physical time-reversal symmetry obeying $T^2=1$, but it can be easily extended also to spinful models with $T^2=-1$ (see Appendix \ref{app:t2}).

Furthemore, motivated by the advances in ultracold atom experiments where synthetic gauge fields may be added and tuned, we discussed the robustness 
of this topological semimetal phase against the breaking of the chiral lattice symmetry, the fluctuations of the flux for realistic system sizes, and the introduction 
of a parabolic trapping potential.

Our results suggest that the role of gauge symmetries in topological phases of matter have been so far underestimated. The tenfold classification of topological insulators and superconductors \cite{ludwig08,hasankane,zhang11} is based on non-spatial  (time-reversal, particle-hole and chiral) symmetries, which are resilient against most sources of disorder. This classification has been extended to spatial symmetries \cite{schnyder15}, including point group symmetries \cite{fu11,zaanen13,diez14} and their magnetic counterparts \cite{mong10,zhang15}. However, gauge symmetries cannot be reduced to them and a systematic study of their effects is still missing.
Also for gapless systems a complete classification of topological phases exists based on non-spatial symmetries \cite{schnyder15,ryu13,zhao13} and only a few generalizations have been considered so far \cite{schnyder14}.

\begin{acknowledgments}
We warmly thank Fabian Hassler and Xi Dai for useful discussions.
 A.T. acknowledges support from the STREP MatterWave. L. L. acknowledges financial support by the ERC-St Grant ColdSIM (No. 307688). M.B. acknowledges support from the EU grant SIQS. I.C.F. thanks the ERC under the EU Seventh Framework Programme (FP7/2007-2013) / ERC Project MUNATOP, the US-Israel Binational Science Foundation, and the Minerva Foundation for support.
\end{acknowledgments}

\vspace{1cm}

\appendix

\section{Alternative gauge choice} \label{app:gauge}

The gauge choice used in the main text is invariant not only under the space-inversion symmetry centered on the lattice sites, but also on space inversion symmetries centered on the lattice cubes, on the plaquettes in the $\hat{y}\hat{z}$ planes and on the links in the $\hat{x}$ direction.
Different gauge choices break some of these symmetries. For example in Ref.~\onlinecite{ketterle14}, Dubcek {\it et al.} use:
\begin{equation} \label{phasesk}
 \theta_x = -\pi\left(x+y\right),\quad \theta_y = 0\,, \quad \theta_z = \pi(x+y).
\end{equation}
In this case $H\left(\{\theta\right\})=H\left(\{-\theta\right\})$, such that $\mathcal{T}=T$. In contrast, the effect of the space inversion centered on the links in the $\hat{x}$ direction is:
\begin{equation}
 e^{i\theta_x} \to -e^{i\theta_x} \,, \quad e^{i\theta_z}\to-e^{i\theta_z},
\end{equation}
such that the Hamiltonian in momentum space is not invariant under the inversion $k_x \to -k_x$ \cite{ketterle14}:
\begin{equation}
 H(k)= -2\omega \left[\sigma_x \cos(k_y) + \sigma_y \sin(k_x) + \sigma_z \cos(k_z) \right].
\end{equation}
In this case a gauge transformation $\mathcal{U}_{xz}=e^{i\pi(x+z)}$ is required to define the physical inversion symmetry: $P=\mathcal{P}\mathcal{U}_{xz}$.
Thus, the $PT$ transformation takes the form $\mathcal{PTU}_{xz}$, which confirms that the system is $PT $invariant but not $\cal{PT} $invariant. 

\section{A generalized model for the $PT$ invariant Weyl semimetal phase} \label{app:general}

The cubic lattice model that we presented in the main text is only a specific example of a $PT$ invariant Weyl semimetal phase. In this section we discuss first an alternative model, based on a stack of brick-wall lattices, that hosts the same gapless topological (Weyl) phase preserving the physical $PT$ invariance; then we show that it is possible to interpolate continuously between the cubic lattice model analyzed in the main text and this alternative lattice.
The alternative model is based on a three-dimensional lattice defined by brick-wall layers piled such that all the lattice sites in different layers, labeled by $z$, overlap when projected on the $\hat{x}\hat{y}$ plane. Therefore all the plaquettes oriented in the $\hat{x}\hat{y}$ planes are rectangular with dimension $2a \times a$, whereas all the plaquettes in the $\hat{x}\hat{z}$ or $\hat{y}\hat{z}$ orientations are square plaquettes. We introduce a magnetic $\pi$-flux in only the vertical plaquettes. In this way, such a three-dimensional lattice can be obtained by suppressing half of the tunnelings along the $\hat{y}$ direction in the cubic lattice model with $\pi$-fluxes in all the directions. The horizontal planes $\hat{x}\hat{y}$, indeed, are constituted only by rectangular plaquettes in which the magnetic flux is doubled to $2\pi$ and, therefore, has no physical effect.

The real space Hamiltonian results:
\begin{multline} 
\label{hambrick}
 H_{BW}=-\omega \sum_{\vec{r} \; : \; x-y={\rm even}} \left[ e^{i\theta_y\left( \vec{r}\right) } c^\dag_{\vec{r}+\hat{y}} \, c_{\vec{r}} + {\rm H.c.} \right] -  \\
 -\omega \sum_{\vec{r}}\left[ e^{i\theta_x\left( \vec{r}\right) } c^\dag_{\vec{r}+\hat{x}} \, c_{\vec{r}} + e^{i\theta_z\left( \vec{r}\right) } c^\dag_{\vec{r}+\hat{z}} \, c_{\vec{r}} + {\rm H.c.} \right] \,
\end{multline}
where we can choose a gauge such that:
\begin{equation} \label{phases}
 \theta_z\left( \vec{r}\right)= \pi (x-y) \,, \quad \theta_x=\theta_y=0\,.
\end{equation}
Considering a two-site unit cell given by the sublattices in the horizontal planes, and labeling it by the pseudospin $\tau_z=\pm 1$, we obtain the following Hamiltonian in momentum space:
\begin{equation}
 \frac{H_{BW}}{\omega} = -2 \tau_z \cos(k_z)  -2 \tau_x \cos(k_x) -  \tau_x \cos(k_y) + \tau_y \sin(k_y) \, .
\end{equation}
The related spectrum results:
\begin{equation}
 E= \pm \omega \sqrt{4\cos^2k_z + (2\cos k_x + \cos k_y)^2 + \sin^2k_y}
\end{equation}
and it is characterized by four Weyl points in $\vec{k}_0^{\pm,\pm}=\left(\pm 2\pi/3,0,\pm \pi/2 \right) $ with a choice of the BZ such that $k_x,k_z \in [-\pi,\pi)$ and $k_y \in [-\pi/2,\pi/2)$.

The previous Hamiltonian is trivially invariant under the canonical time-reversal symmetry $\mathcal{T}$, since the Hamiltonian \eqref{hambrick} has only real hopping amplitudes. Therefore $\mathcal{T}=T$. Besides, the lattice is invariant under a space inversion centered in one of the rectangular plaquettes, such that $\mathcal{P}$ is described by:
\begin{equation}
\label{canonicalP}
 x \to -x \,, \quad y\to 1-y \,, \quad z \to -z.
\end{equation}
However, our gauge choice is such that $\theta_z$ is not invariant under this canonical transformation and $e^{i\theta_z} \xrightarrow{\mathcal{P}} -e^{i\theta_z}$. Therefore the physical space inversion must be defined by $P=\mathcal{P}\mathcal{U}_z$, where the gauge transformation $\mathcal{U}_z c_{\vec{r}} = e^{i \pi z}c_{\vec{r}}$ enters.

It is easy to see that $T$ maps $\vec{k}_0^{a,b} \to \vec{k}_0^{-a,-b}$ whereas $P=\mathcal{P}\mathcal{U}_z$ gives $\vec{k}_0^{a,b} \to \vec{k}_0^{-a,b}$.
Therefore also in this case, despite the $PT$ invariance, Weyl points with different chiralities have different positions in momentum space, thus ensuring the possibility of the Weyl phase.

It is possible to interpolate between the brick wall lattice model in Eq. \eqref{hambrick} and the cubic lattice model with $\pi$-fluxes discussed in the main text. This is achieved by adding the missing hopping terms along the odd links in the $\hat{y}$ directions of the first Hamiltonian. More formally, we define:
\begin{equation}
 H(\alpha)= H_{BW} - \alpha \, \omega \sum_{\vec{r} \;{\rm s.t.} \; x-y={\rm odd}} \left[ e^{i\theta_y\left( \vec{r}\right) } c^\dag_{\vec{r}+\hat{y}}c_{\vec{r}} + {\rm H.c.} \right] \, .
 \label{halpha}
\end{equation}
We complete the definition of the phases in Eq. \eqref{phases} with:
\begin{equation}
 \theta_y\left( \vec{r}\right) = \pi (x-y) \, .
\end{equation}
With this choice of the gauge, $H(\alpha)$ describes again a cubic lattice model with $\pi$-fluxes in all the plaquettes, as in the main text. It interpolates between $H(\alpha=0)=H_{BW}$ and the Hamiltonian $H(\alpha=1)$ which is analogous, up a rotation, to the cubic lattice model with the hoppings in Eq. \eqref{phasesk}. For each value of $\alpha$ the canonical $\mathcal{T}$ is preserved, whereas the physical inversion symmetry $P$ coincides with the one specified above for $H_{BW} $. 

The position along $k_x$ of the Weyl points varies with $\alpha$, $\vec{k}_0^{\pm,\pm}=\left(\pm q(\alpha),0,\pm \pi/2 \right)$, but the relation  $P\ket{\vec{k}_0^{a,b}}= \ket{\vec{k}_0^{-a,b}}$ remains valid for each value of $\alpha \in [0,1]$, such that all the Hamiltonians obtained in Eq. (\ref{halpha}) describe a $PT$ invariant Weyl semimetal.

\section{Kernel polynomial method} \label{app:kernel}

In Fig. 3 of the main text, the density of states $\rho(E)$ is computed using the kernel polynomial method \cite{Weisse06}. In the following we summarize the basic steps of this approximation. For a Hamiltonian $H$ with eigenvalues $E_n$, the density of states is given by

\begin{equation}\label{eq:dos}
 \rho(E) = \sum_{n} \delta(E - E_n).
\end{equation}

We will expand Eq.~\eqref{eq:dos} in a series of Chebyshev polynomials, $T_n(x) = \cos[n\, {\rm arccos}(x)]$, which obey the recursion relations
\begin{equation}
 T_{n+1}(x) = 2x \, T_n(x) - T_{n-1}(x),
\end{equation}
with $T_0(x)=1$ and $T_1(x)=x$. Since the polynomials are defined only in the interval $[-1,1]$, the first step is to rescale the Hamiltonian such that its spectrum is contained this interval:
\begin{equation}
\widetilde{H} = \frac{H - b}{a},
\end{equation}
such that the energy becomes
\begin{equation}
\widetilde{E} = \frac{E - b}{a}.
\end{equation}
Setting $\omega=1$ in the Hamiltonian (2) of the main text, we choose the values $b=0$ and $a=8$, for which the rescaled energies $\widetilde{E} \in [-1,1]$. Following Ref.~\onlinecite{Weisse06}, the rescaled density of states, $\widetilde{\rho}(\widetilde{E}) = \sum_{n} \delta(\widetilde{E} - \widetilde{E}_n)$,
can be written as an infinite series
\begin{equation}\label{eq:expansion}
 \widetilde{\rho}(\widetilde{E}) = \frac{1}{\pi\sqrt{1-\widetilde{E}^2}}\left[ \mu_0 +2\sum_{n=1}^\infty \mu_n T_n(\widetilde{E}) \right],
\end{equation}
with the expansion coefficients given by 
\begin{equation}\label{eq:coeff}
 \mu_n = {\rm Tr}\, T_n (\widetilde{H}),
\end{equation}
where ${\rm Tr}$ denotes the trace.

For our numerical purposes, we truncate the expansion of Eq.~\eqref{eq:expansion} by keeping only the first $N$ coefficients:
\begin{equation}\label{eq:truncation}
 \widetilde{\rho}(\widetilde{E}) \simeq \frac{1}{\pi\sqrt{1-\widetilde{E}^2}}\left[ \mu_0 +2\sum_{n=1}^{N-1} \mu_n T_n(\widetilde{E}) \right].
\end{equation}
This approximation leads to fluctuations in the density of states, also known as Gibbs oscillations, which we damp by modifying the expansion coefficients $\mu_n\to g_n\mu_n$. For the density of states plotted in Fig. 3 we have used the so-called Jackson kernel, corresponding to
\begin{equation}
 g_n = \frac{(N-n+1)\cos\displaystyle{\frac{\pi n}{N+1}}+\sin{\frac{\pi n}{N+1}}\cot{\frac{\pi}{N+1}}}{N+1},
\end{equation}
and computed the first $N=2048$ terms of Eq.~\eqref{eq:truncation}.

Finally, to determine the coefficients \eqref{eq:coeff} we have used a stochastic evaluation of the trace
\begin{equation}\label{eq:coeffrnd}
 \mu_n = {\rm Tr}\, T_n (\widetilde{H}) \simeq \frac{1}{R}\sum_{r=0}^{R-1} \langle r | T_n(\widetilde{H}) | r \rangle,
\end{equation}
where $|r\rangle$ are random states with entries drawn from the Gaussian distribution with zero mean and unit variance ${\cal N}(0, 1)$. Each coefficient was computed using $R=10$ random vectors.

\section{Spin $1/2$ $PT$-invariant Weyl semimetal} \label{app:t2}

It is possible to obtain a $PT$-invariant Weyl semimetal phase also in spinful models characterized by $T^2=-1$, differently from the spinless case of Eq. \eqref{peierls1a} for which $T^2=+1$. To investigate this scenario, let us study a toy model of spin $1/2$ fermions on the cubic lattice in which both the spin species are subject to $\pi$-fluxes in all the plaquettes and they are mixed by non-trivial tunneling operators. We consider the Hamiltonian:
\begin{equation}
H_2(\{\theta\}) = -  \omega \, 
\sum_{\vec{r}  ,  \hat{j},s,s'}   c^{\dagger}_{\vec{r} + \hat{j},s} O_{ss'}
e^{i \theta_{j}\left(\vec{r}\right)} 
 c_{\vec{r},s' } + \ \mathrm{H.c.} \, ,
\label{peierls2a}
\end{equation}
where the phases $\theta_{j}$ are the ones defined in Eq. \eqref{eq:abthetas} and we introduced a generic $2\times2$ unitary tunneling operator of the form $O_{ss'}=e^{i\vec{\alpha}\vec{\sigma}}$. Here $\vec{\alpha}$ is a generic position-independent vector and $\vec{\sigma}$ is the vector of the Pauli matrices. 
In this case the canonical time-reversal transformation must be defined as $\mathcal{T}=\sigma_y\mathcal{K}$, where $\mathcal{K}$ is the complex conjugation. It is easy to see that $\mathcal{T}O\mathcal{T}^\dag=O$, therefore, once again, we obtain $\mathcal{T} H_2(\{\theta\}) \mathcal{T}^\dag=H(\{-{\theta}\})$. Analogously to the spinless case, Eq. \eqref{Tphys} holds. Hence, also for $H_2$, the physical time-reversal symmetry reads $T=\mathcal{U}_y\mathcal{T}$, but, differently from the spinless case, we have $T^2=-1$.

Concerning the inversion symmetry, beside mapping $\vec{r}$ into $-\vec{r}$, the parity must flip $\vec{\alpha}\vec{\sigma}$ into $-\vec{\alpha}\vec{\sigma}$. Therefore $P=\mathcal{P}=i\hat{\beta}\vec{\sigma} \tilde{\mathcal{P}}$ where $\hat{\beta}$ is a unit vector orthogonal to $\vec{\alpha}$ and $\tilde{\mathcal{P}}$ describes the coordinate transformation \eqref{inversion} corresponding to $\mathcal{P}$ in the spinless case. In this way $\mathcal{P}O\mathcal{P}^\dag=O^\dag$ such that $\mathcal{P}H_2\mathcal{P}^\dag=H_2$. Also in this case $[T,P]=0$.

The Hamiltonian \eqref{peierls2a} describes a Weyl semimetal with eight Weyl points in the Brillouin zone. With respect to the spinless case, the Weyl points are displaced in momentum space by the two vectors $\pm \left(|\vec{\alpha}|,|\vec{\alpha}|,|\vec{\alpha}| \right)$, such that $\vec{k_0}^{\pm,\pm,s}=\left(\pm \pi/2 +s |\vec{\alpha}|,\pm \pi/2 +s |\vec{\alpha}|,\pi/2 +s |\vec{\alpha}| \right)$, where $s=\pm 1$ describes the two eigenstates of $\vec{\alpha}\vec{\sigma}$. For $|\vec{\alpha}| \neq 0, \pi/2$ this implies that the Weyl points do not overlap in momentum space and Eq. \eqref{peierls2a} defines a spin 1/2 $PT$-invariant Weyl semimetal with $T^2=-1$.

The transformation of the Weyl points under $PT$ can be easily obtained by generalizing the spinless case and considering that $P$ maps $s$ into $-s$.


\begin{thebibliography}{99}
\bibitem{notetras}
The magnetic unit cell is found imposing the commutation of translations along the different spatial directions. These translations must involve, in the presence of commensurate magnetic fluxes, multiples of the lattice constant \cite{landau}.
\end{thebibliography}
\end{document}